\documentclass[amstex,twocolumn,showpacs,floats,floatfix,superscriptaddress,aps,pra]{revtex4}
\usepackage{amssymb}
\usepackage{amsmath}
\usepackage{calc}
\usepackage{graphicx}
\usepackage{bm}

\input{tcilatex}

\begin{document}

\author{G. S. Vasilev}
\affiliation{Department of Physics, Sofia University, James Bourchier 5 blvd, 1164 Sofia,
Bulgaria}
\affiliation{CloudRisk Ltd}
\title{Time-dependent Heston model.}
\date{\today }

\begin{abstract}
This work presents an exact solution to the generalized Heston model, where
the model parameters are assumed to have linear time dependence The solution
for the model in expressed in terms of confluent hypergeometric functions.
\end{abstract}

\maketitle


\section{Introduction}


The paper presents generalization of Heston's (1993) \cite{Heston}
stochastic volatility model based on time-dependent model parameters. The
Heston model is one of the most widely used stochastic volatility (SV)
models today. The Heston model can be viewed as a stochastic volatility
generalization of the Black and Scholes (1973) \cite{BS} model and includes
it as a special case. It is well known that the implied volatility smile
used with Black-Scholes formula tends to systematically misprice
out-of-the-money and in-the-money options if the volatility implied from the
at-the-money option has been used. According to this effect various SV
option pricing models (Hull and White (1987) \cite{HullWhite}, Stein and
Stein (1991) \cite{Stein}, Heston (1993) \cite{Heston}) have been developed
to capture the \textquotedblleft smile\textquotedblright\ effect. The
popularity and the attractiveness of the Heston model lies in the
combination of its three main features: it does not allow negative
volatility, it allows the correlation between asset returns and volatility
and it has a closed-form pricing formula. However, to use Heston model for
option pricing one needs to know the model structural parameters. This
brings us to the calibration problem and in the case of Heston model it
faces several difficulties. For example the data used for model calibration
are observed at discrete times, but the model is built under a
continuous-time framework.

Within the calibration problem lies the need of generalization of the Heston
model. Generally speaking, because the prices from stochastic engines (in
our case the Heston model) are not supported by market prices, as a result
the model parameters have to be recalibrated every day to new market data.
This solution to calibration issue, apart from time consuming is not
consistent with an accurate description of the dynamics. This is the main
motivation to consider the Heston model with time-dependent parameters.


\section{Heston's stochastic volatility model}


In this section the overview of the Heston model will be presented. We will
derive the pricing partial differential equation (PDE), which forms the
basis of the derivation of the characteristic function in the next section.
The Heston model was first introduced in \cite{Heston}, where we have two
stochastic differential equations, one for the underlying asset price $S(t)$
and one for the variance $V(t)$ of log$S(t$):%
\begin{eqnarray}
\frac{dS(t)}{S(t)} &=&\mu (t)dt+\sqrt{V(t)}dW_{1}  \label{Heston} \\
dV(t) &=&\kappa \left( \theta -V(t)\right) dt+\eta \sqrt{V(t)}dW_{2}.  \notag
\end{eqnarray}

Here $\kappa \geq 0,\theta $ $\geq 0$ and $\eta >0$ stands for speed of mean
reversion, the mean level of variance and the volatility of the volatility,
respectively. Furthermore, the Brownian motions $W_{1}$ and $W_{2}$ are
assumed to be correlated with correlation coefficient $\rho .$ From the
Heston model definition given by Eq.(\ref{Heston}), it is trivial
observation that SDE for the variance can be recognized as a mean-reverting
square root process - CIR process, a process originally proposed by Cox,
Ingersoll \& Ross (1985) \cite{CIR} to model the spot interest rate. Using
Ito's lemma and standard arbitrage arguments we arrive at Garman's pricing
PDE for the Heston model which reads

\begin{eqnarray}
&&\frac{\partial P}{\partial t}+\frac{1}{2}VS^{2}\frac{\partial ^{2}P}{%
\partial S^{2}}+\rho \eta SV\frac{\partial ^{2}P}{\partial S\partial V}+%
\frac{1}{2}V\eta ^{2}\frac{\partial ^{2}P}{\partial V^{2}}+  \notag \\
S\frac{\partial P}{\partial S}-rP &=&\left[ \kappa \left( V-\theta \right)
-\lambda V\right] \frac{\partial P}{\partial V},
\end{eqnarray}

where $\lambda $ is the market price of volatility risk. For the reader
convenience a detail derivation is given in Appendix A. We further simplify
the pricing PDE given by Eq.(\ref{Heston-PDE}) by defining the forward
option price
\begin{equation*}
C_{u}\left( x(t),V(t),t\right) =e^{r(T-t)}P\left( S(t),V(t),t\right)
\end{equation*}%
in which%
\begin{equation*}
x(t)=\log \left[ \frac{e^{r(T-t)}S(t)}{K}\right]
\end{equation*}

Finally, define $\tau =T-t,$ then we obtain the so-called forward equation
in the form%
\begin{eqnarray}
&&-\frac{\partial C_{u}}{\partial t}+\frac{1}{2}V\left[ \frac{\partial
^{2}C_{u}}{\partial x^{2}}-\frac{\partial C_{u}}{\partial x}\right] +
\label{Heston-forwardPDE} \\
\rho \eta V\frac{\partial ^{2}C_{u}}{\partial x\partial V}+\frac{1}{2}V\eta
^{2}\frac{\partial ^{2}C_{u}}{\partial V^{2}} &=&\kappa \left( V-\theta
\right) \frac{\partial C_{u}}{\partial V}  \notag
\end{eqnarray}


\subsection{Characteristic function of the Heston model}


Prior presenting the results related to characteristic function of the
Heston model, we will refer to some results for affine diffusion processes.
Following the work of Duffie, Pan and Singleton (2000) \cite%
{Duffie-Pan-Singleton}, for affine diffusion processes the characteristic
function of $x(T)$ reads%
\begin{equation}
f(x,V,\tau ,\omega )=\exp \left[ A\left( \omega ,\tau \right) +B\left(
\omega ,\tau \right) V+C\left( \omega ,\tau \right) x\right]
\label{characteristic function}
\end{equation}%
where is assumed $V\equiv V(t)$ and $x\equiv x(t).$Moreover, the
characteristic function must satisfy the following initial condition%
\begin{equation*}
f(x,V,0,\omega )=\exp \left[ i\omega x(T)\right] ,
\end{equation*}

which implies that
\begin{equation}
A\left( \omega ,0\right) =0;\text{ \ \ }B\left( \omega ,0\right) =0;\text{ \
}C\left( \omega ,0\right) =i\omega  \label{initial conditions}
\end{equation}

Substituting characteristic function given by Eq.(\ref{characteristic
function}) in forward pricing PDE Eq.(\ref{Heston-forwardPDE}), according to
the initial conditions, can be shown that the characteristic function
simplifies to%
\begin{equation}
f(x,V,\tau ,\omega )=\exp \left[ A\left( \omega ,\tau \right) +B\left(
\omega ,\tau \right) V+i\omega x\right]  \label{characteristic function-1}
\end{equation}

The functions $A\left( \omega ,\tau \right) $ and $\ B\left( \omega ,\tau
\right) $ satisfy the following system of ordinary differential equations
ODE
\begin{subequations}
\label{A-B-ODE}
\begin{eqnarray}
\frac{dA}{d\tau } &=&aB,\text{ \ \ }A\left( \omega ,0\right) =0  \label{A-B1}
\\
\frac{dB}{d\tau } &=&\alpha -\beta B+\gamma B^{2},\text{ \ \ }B\left( \omega
,0\right) =0,  \label{A-B2}
\end{eqnarray}

where
\end{subequations}
\begin{eqnarray}
a &=&\kappa \theta  \label{A-B-ODE-params} \\
\alpha &=&-\frac{1}{2}\left( \omega ^{2}+i\omega \right)  \notag \\
\beta &=&\kappa -\rho \eta i\omega  \notag \\
\gamma &=&\frac{1}{2}\eta ^{2}  \notag
\end{eqnarray}

and $\omega \in \mathbf{R.}$

For the reader's convenience, in Appendix B we will give the derivation of
the ODE\ in Eq.(\ref{A-B-ODE}).


\section{Heston's stochastic volatility model with time dependent model
parameters}


Since
\begin{equation*}
f(x,V,\tau ,\omega )=\mathbf{E}^{\mathbf{Q}}\left[ e^{i\omega x(T)}\right] ,
\end{equation*}

where $\mathbf{Q}$ is some risk-neutral measure, the problem for exact
analytic solution for the Heston model is cast to problem for the respective
characteristic function or more precisely solutions for the $A\left( \omega
,\tau \right) $ and $\ B\left( \omega ,\tau \right) $ factors for the
characteristic function. There are several directions towards
generalizations of the Heston model with time-dependent parameters.
Mikhailov and Nogel (2003) \cite{Mikhailov} indicate that $\theta $ can be
relaxed from the constrain to be constant. As can be seen from Eq.(\ref{A-B1}%
) the parameter $\theta $ can be assumed time dependent and the exact
solution of $A\left( \omega ,\tau \right) $ is still possible. Mikhailov and
Nogel indicate that other choices of time dependent parameters are still
possible but the general solution of the ODE system Eq.(\ref{A-B-ODE}) is
restricted from the Riccati equation given in Eq.(\ref{A-B2}). Other
possible generalizations are related to time discrimination - piece-wise
constant parameters or asymptotic solutions.

Our approach assumes that all model parameters have linear time dependence.
I.e. $\kappa ,$ $\theta ,\eta $ and $\rho $ are linear in respect in time.
Although this is a restrictive, we would like to note that locally,
arbitrary time dependence is reduced to linear. Also such generalization is
the simplest non-trivial, where all model parameters are time-dependent.

We should stress that our main concern is finding a solution of the Eq.(\ref%
{A-B2}), because the solution of the $A\left( \omega ,\tau \right) $ is
simply given by integration with integrand the desired solution of $B\left(
\omega ,\tau \right) $ times $a.$


\subsection{Constant volatility of the volatility parameter}


We should stress that even within this framework, assuming linear
time-dependence for all model parameters, the solution is complicated. With
this reasoning, we first will consider constant $\eta $ and linear $\kappa ,$
$\theta $ and $\rho $.

\begin{eqnarray}
\eta &=&const  \label{const-1} \\
\kappa &=&\kappa _{1}\tau +\kappa _{2}  \notag \\
\theta &=&\theta _{1}\tau +\theta _{2}  \notag \\
\rho &=&\rho _{1}\tau +\rho _{2}  \notag
\end{eqnarray}

Having defined model parameters the next step is looking for solutions.
Because the analytic theory for second order linear ODE\ with time-dependent
coefficients is well studied, our first step is to cast the Riccati equation
in Eq.(\ref{A-B2}) to linear second order ODE. In Eq.(\ref{A-B2}), after the
substitution
\begin{equation}
B=-\frac{\dot{D}}{\gamma D}  \label{B-D-transform}
\end{equation}

where $\dot{D}=\frac{dD}{d\tau }$, we obtain%
\begin{equation}
\frac{d^{2}D}{d\tau ^{2}}+\beta \frac{dD}{d\tau }+\alpha \gamma D=0
\label{Lin-eq-D}
\end{equation}

According to the substitution Eq.(\ref{B-D-transform}) and the initial
conditions Eq.(\ref{initial conditions}) the new initial condition for $%
D(\tau )$ reads
\begin{subequations}
\label{initail-D}
\begin{equation}
\dot{D}(0)=0  \label{D2}
\end{equation}

Note that we should have two initial conditions as we intend to solve second
order ODE, but because the prime equation is Eq.(\ref{A-B2}), which is first
order, the initial condition in Eq.(\ref{D2}) is sufficient. With the
notation used we obtain the following equation to solve

\end{subequations}
\begin{equation}
\frac{d^{2}D}{d\tau ^{2}}+(g_{1}\tau +g_{2})\frac{dD}{d\tau }+\alpha \gamma
D=0,  \label{D-confluent-hyper}
\end{equation}

where
\begin{eqnarray*}
g_{1} &=&\kappa _{1}-i\rho _{1}\eta \omega \\
g_{2} &=&\kappa _{2}-i\rho _{2}\eta \omega
\end{eqnarray*}

Eq.(\ref{D-confluent-hyper}) can be transformed to confluent hypergeometric
equation via the substitution%
\begin{equation}
z=-\frac{(g_{1}\tau +g_{2})^{2}}{2g_{1}}  \label{hypergeometric-sub}
\end{equation}

After simple algebra we obtain%
\begin{equation}
z\frac{d^{2}D}{dz^{2}}+\left( \frac{1}{2}-z\right) \frac{dD}{dz}-\frac{%
\alpha \gamma }{2g_{1}}D=0  \label{D-confluent-hyper-finall}
\end{equation}

If we compare the equation given Eq.(\ref{D-confluent-hyper-finall}) with
the standard form of the confluent hypergeometric equation, which reads

\begin{equation}
z\frac{d^{2}w}{dz^{2}}+\left( b-z\right) \frac{dw}{dz}-aw=0
\label{conf-hypergeometric}
\end{equation}

we immediately can write the desired solution. Basic reference formulas
about confluent hypergeometric equation and its solutions are given in
Appendix C. The solution for Eq.(\ref{D-confluent-hyper-finall}) is
\begin{eqnarray}
D(\tau ) &=&A_{1}M\left[ \frac{\alpha \gamma }{2g_{1}},\frac{1}{2},-\frac{%
(g_{1}\tau +g_{2})^{2}}{2g_{1}}\right] +  \label{D-solution} \\
&&A_{2}U\left[ \frac{\alpha \gamma }{2g_{1}},\frac{1}{2},-\frac{(g_{1}\tau
+g_{2})^{2}}{2g_{1}}\right] .
\end{eqnarray}

Here $M\left[ \frac{\alpha \gamma }{2g_{1}},\frac{1}{2},-\frac{(g_{1}\tau
+g_{2})^{2}}{2a}\right] $ and $U\left[ \frac{\alpha \gamma }{2g_{1}},\frac{1%
}{2},-\frac{(g_{1}\tau +g_{2})^{2}}{2a}\right] $ are standard notations for
the Kummer functions \cite{AS} representing the two linear independent
solutions of Eq.(\ref{conf-hypergeometric}). From Appendix C, because $b=%
\frac{1}{2},$ second solution $U(a,b,z)$ of Eq.(\ref{conf-hypergeometric})
is given by convergent series for all values if $a$ and $b$. Next step
towards completing the solution for the time-dependent Heston model is to
obtain the integration constants $A_{1}$ and $A_{2}$ from the initial
condition.

Using Eq.(\ref{dz-M}) and Eq.(\ref{dz-U}) form the solution given in Eq.(\ref%
{D-solution}) and the initial condition in Eq.(\ref{D2}), we obtain

\begin{widetext}
\begin{equation}
\left. \frac{d}{d\tau }D(\tau )\right\vert _{\tau =0}=A_{1}\frac{\alpha
\gamma }{g_{1}}M\left[ \frac{\alpha \gamma }{2g_{1}}+1,\frac{3}{2},-\frac{%
(g_{1}\tau +g_{2})^{2}}{2g_{1}}\right] (-g_{1}\tau -g_{2})\left.
+A_{2}\left( -\frac{\alpha \gamma }{2g_{1}}\right) U\left[ \frac{\alpha
\gamma }{2g_{1}}+1,\frac{3}{2},-\frac{(g_{1}\tau +g_{2})^{2}}{2g_{1}}\right]
(-g_{1}\tau -g_{2})\right\vert _{\tau =0}  \notag
\end{equation}
\end{widetext}

From the above Eq.(\ref{dz-D-initial-con}) we obtain the functional relation
between $A_{1}$ and $A_{2}$

\begin{equation}
A_{1}=\frac{A_{2}}{2}\frac{U\left[ (\alpha \gamma
+2g_{1})/2g_{1},3/2,-(g_{2})^{2}/2g_{1}\right] }{M\left[ (\alpha \gamma
+2g_{1})/2g_{1},3/2,-(g_{2})^{2}/2g_{1}\right] }=\frac{A_{2}}{2}F
\label{A1-A2}
\end{equation}

To shorten the notations we have used $F$ to denote the $U/M$ fraction.

As we have mentioned due to the homogenous transformation given in Eq.(\ref%
{B-D-transform}) the functional relation between $A_{1}$ and $A_{2}$ in Eq.(%
\ref{A1-A2}) is sufficient to determine the function $\ B\left( \omega ,\tau
\right) $ of the characteristic function for the Heston model

\begin{widetext}
\begin{equation}
B\left( \omega ,\tau \right) =\frac{\alpha }{g_{1}}(g_{1}\tau +g_{2})\left\{
\frac{U\left[ \frac{\alpha \gamma }{2g_{1}}+1,\frac{3}{2},-\frac{(g_{1}\tau
+g_{2})^{2}}{2g_{1}}\right] -F\text{ }M\left[ \frac{\alpha \gamma }{2g_{1}}%
+1,\frac{3}{2},-\frac{(g_{1}\tau +g_{2})^{2}}{2g_{1}}\right] }{F\text{ }M%
\left[ \frac{\alpha \gamma }{2g_{1}},\frac{1}{2},-\frac{(g_{1}\tau
+g_{2})^{2}}{2g_{1}}\right] +U\left[ \frac{\alpha \gamma }{2g_{1}},\frac{1}{2%
},-\frac{(g_{1}\tau +g_{2})^{2}}{2g_{1}}\right] }\right\}
\end{equation}
\end{widetext}


\subsection{Linear in time volatility of the volatility parameter}


Let us derive the solution for the general case of linear time dependence.
The model parameters have the following form

\begin{eqnarray}
\eta &=&\eta _{1}\tau +\eta _{2}  \label{const-2} \\
\kappa &=&\kappa _{1}\tau +\kappa _{2}  \notag \\
\theta &=&\theta _{1}\tau +\theta _{2}  \notag \\
\rho &=&\rho _{1}\tau +\rho _{2}  \notag
\end{eqnarray}

Again one can apply the transformation given by Eq.(\ref{B-D-transform}) to
the Eq.(\ref{A-B2}). After some algebra we obtain the following equation

\begin{equation}
\frac{d^{2}D}{d\tau ^{2}}+(h_{1}\tau ^{2}+h_{2}\tau +h_{3})\frac{dD}{d\tau }%
+(k_{1}\tau +k_{2})^{2}D=0,  \label{D-HeunT}
\end{equation}

where the coefficients $h_{1},h_{2},h_{3},k_{1}$ and $k_{2}$ are given as
follows%
\begin{eqnarray*}
h_{1} &=&-i\omega \rho _{1}\eta _{1} \\
h_{2} &=&\kappa _{1}-(\rho _{1}\eta _{2}+\rho _{2}\eta _{1})i\omega \\
h_{3} &=&\kappa _{2}-\rho _{2}\eta _{2}i\omega \\
k_{1} &=&\frac{\eta _{1}\sqrt{\alpha }}{\sqrt{2}} \\
k_{2} &=&\frac{\eta _{2}\sqrt{\alpha }}{\sqrt{2}}
\end{eqnarray*}

The solution of the Eq.(\ref{D-HeunT}) can be cast to the tree-confluent
Heun equation \cite{Ronveaux} given by Eq.(\ref{Triconfluent-eq1}).
\begin{eqnarray}
D(\tau ) &=&A_{1}\exp \left( f\right) \text{ HeunT}(\alpha ,\beta ,\gamma
,z)+  \label{D-solution-HeunT} \\
&&A_{2}\exp \left[ -\frac{\tau \left( k_{1}\right) ^{2}}{h_{1}}\right] \text{%
HeunT}(\alpha ,\beta ,\gamma ,z)
\end{eqnarray}

where we have used the following notations%
\begin{equation}
f=-\frac{\tau \left[ 2\left( h_{1}\tau \right) ^{2}+3h_{1}h_{2}\tau
+6h_{1}h_{3}-6\left( k_{1}\right) ^{2}\right] }{6h_{1}}  \label{f}
\end{equation}

\begin{eqnarray}
\alpha &=&\frac{3^{2/3}}{2\left( h_{1}\right) ^{8/3}}\left[ \left(
h_{2}k_{1}\right) ^{2}+2\left( k_{2}k_{1}\right)
^{2}-2h_{1}h_{2}k_{2}k_{1}\right.  \label{alfa} \\
&&\left. -2\left( k_{1}\right) ^{2}h_{1}h_{3}+2\left( k_{1}\right) ^{4}
\right]
\end{eqnarray}

\begin{equation}
\beta =-\frac{3\left[ \left( k_{1}\right) ^{2}h_{2}+\left( k_{1}\right)
^{2}-2h_{1}k_{2}k_{1}\right] }{\left( h_{1}\right) ^{2}}  \label{beta}
\end{equation}

\begin{equation}
\gamma =\frac{3^{1/3}\left[ 4h_{1}h_{3}-\left( h_{2}\right) ^{2}-8\left(
k_{1}\right) ^{2}\right] }{4\left( h_{1}\right) ^{4/3}}  \label{gamma}
\end{equation}

Again for reader convenience basic definitions and formulas for Heun
equations are given in Appendix D.

\begin{figure}[tb]
\includegraphics[width=75mm]{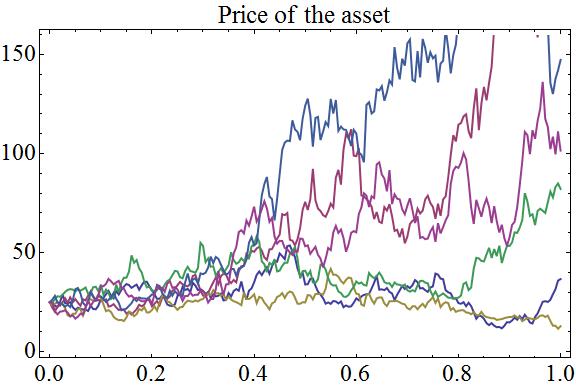}
\caption{Asset price govern by the Heston stochastic proces .}
\label{asset}
\end{figure}

\begin{figure}[tb]
\includegraphics[width=75mm]{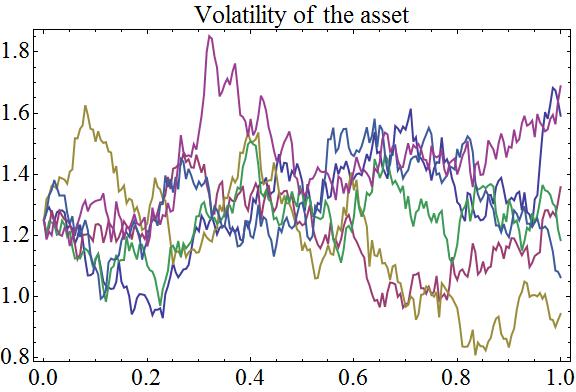}
\caption{Volatility related to Heston stochastic prosec.}
\label{vol}
\end{figure}


\section{CONCLUSIONS}


The Heston model is among the most popular stochastic volatility models due
to its analytical tractability. Nevertheless, the complete use of the Heston
model is still challenging because it has a closed formula only when the
parameters are constant or piecewise constant or within the asymptotic
limits. The aim of this research is to fill this theoretical gap and to
enrich the usability of this very popular model in respect to calibration
issues.

There are several concluding remarks that should be also noted:

A similar to the presented above approach is applicable to the SVJ models
and more precisely to the Bates model. This problem will be addressed in
more details in forthcoming research.

Within the presented work we have assumed that all model parameters obey
linear time dependence. This assumption have been made for consistency, but
the more general assumption allow arbitrary time dependence for the $\theta $
Heston parameter.

\acknowledgments This work has been supported by the project QUANTNET -
European Reintegration Grant (ERG) - PERG07-GA-2010-268432.

\section{Appendix A}


This appendix summarize a key formulas and their derivation for the Heston
model. Using It\^{o}'s lemma for a function $V(S_{1},S_{2},t)$, which is
twice differentiable with respect to $S_{1}$ and $S_{2}$, and once with
respect to $t$ , where

\begin{eqnarray*}
dS_{1} &=&a_{1}(S_{1},S_{2},t)dt+b_{1}(S_{1},S_{2},t)dW_{1} \\
dS_{2} &=&a_{2}(S_{1},S_{2},t)dt+b_{2}(S_{1},S_{2},t)dW_{2}
\end{eqnarray*}

and $W_{1}$ and $W_{2}$ are correlated with correlation coefficient $\rho ,$
for the differential of $V(S_{1},S_{2},t)$ due to It\^{o}'s lemma we obtain
\begin{widetext}
\begin{equation}
dV=\frac{\partial V}{\partial t}dt+\frac{\partial V}{\partial S_{1}}dS_{1}+%
\frac{\partial V}{\partial S_{2}}dS_{2}+\frac{1}{2}b_{1}^{2}\frac{\partial
^{2}V}{\partial S_{1}^{2}}dt++\frac{1}{2}b_{2}^{2}\frac{\partial ^{2}V}{%
\partial S_{2}^{2}}dt+\rho b_{1}b_{2}\frac{\partial ^{2}V}{\partial
S_{1}\partial S_{2}}  \notag
\end{equation}
\end{widetext}
It is standard to use the above equation Eq.(\ref{V-Ito}) for a function
assuming a self-financing portfolio with value consisting of an option with
value $P(S,V,t)$, $-\Delta $ units of the

underlying asset $S$ and, in order to hedge the risk associated with the
random volatility, $-\Delta _{1}$ units of another option with value $%
P_{1}(S,V,t).$ Hence we get%
\begin{equation*}
\Pi =P-\Delta S-\Delta _{1}P_{1}
\end{equation*}

Using again It\^{o}'s lemma after some algebra we obtain%
\begin{widetext}
\begin{eqnarray*}
d\Pi  &=&\left( \frac{\partial P}{\partial t}+\frac{1}{2}VS^{2}\frac{%
\partial ^{2}P}{\partial S^{2}}+\rho \eta SV\frac{\partial ^{2}P}{\partial
S\partial V}+\frac{1}{2}V\eta ^{2}\frac{\partial ^{2}P}{\partial V^{2}}%
\right) dt-\Delta _{1}\left( \frac{\partial P_{1}}{\partial t}+\frac{1}{2}%
VS^{2}\frac{\partial ^{2}P_{1}}{\partial S^{2}}+\rho \eta SV\frac{\partial
^{2}P_{1}}{\partial S\partial V}+\frac{1}{2}V\eta ^{2}\frac{\partial
^{2}P_{1}}{\partial V^{2}}\right) dt \\
&&+\left( \frac{\partial P}{\partial S}-\Delta _{1}\frac{\partial P_{1}}{%
\partial S}-\Delta \right) dS+\left( \frac{\partial P}{\partial V}-\Delta
_{1}\frac{\partial P_{1}}{\partial V}\right) dV
\end{eqnarray*}
\end{widetext}

Following the standard approach if one assumes
\begin{subequations}
\begin{eqnarray}
\frac{\partial P}{\partial S}-\Delta _{1}\frac{\partial P_{1}}{\partial S}%
-\Delta  &=&0  \label{arbitrage conditions} \\
\frac{\partial P}{\partial V}-\Delta _{1}\frac{\partial P_{1}}{\partial V}
&=&0  \notag
\end{eqnarray}%
\end{subequations}
one can get a risk-free portfolio. The standard approach requires
to eliminate arbitrage opportunities and hence the return of this
risk-free
portfolio must equal the (deterministic) risk-free rate of return $r:$%
\begin{widetext}
\begin{eqnarray}
d\Pi  &=&\left( \frac{\partial P}{\partial t}+\frac{1}{2}VS^{2}\frac{%
\partial ^{2}P}{\partial S^{2}}+\rho \eta SV\frac{\partial ^{2}P}{\partial
S\partial V}+\frac{1}{2}V\eta ^{2}\frac{\partial ^{2}P}{\partial V^{2}}%
\right) dt-\Delta _{1}\left( \frac{\partial P_{1}}{\partial t}+\frac{1}{2}%
VS^{2}\frac{\partial ^{2}P_{1}}{\partial S^{2}}+\rho \eta SV\frac{\partial
^{2}P_{1}}{\partial S\partial V}+\frac{1}{2}V\eta ^{2}\frac{\partial
^{2}P_{1}}{\partial V^{2}}\right) dt  \notag \\
&=&r\Pi dt=r\left( P-\Delta S-\Delta _{1}P_{1}\right) dt  \notag
\end{eqnarray}
\end{widetext}

Using the conditions in Eq.(\ref{arbitrage conditions}) the above
equation for the risk-free porfolio Eq.(\ref{P-eq}) become

\begin{eqnarray*}
&&\left( \frac{\partial P}{\partial t}+\frac{1}{2}VS^{2}\frac{\partial ^{2}P%
}{\partial S^{2}}+\rho \eta SV\frac{\partial ^{2}P}{\partial S\partial V}%
\right.  \\
&&\left. +\frac{1}{2}V\eta ^{2}\frac{\partial ^{2}P}{\partial V^{2}}+rS\frac{%
\partial P}{\partial S}-rV\right) /\frac{\partial P}{\partial V} \\
&=&\left( \frac{\partial P_{1}}{\partial t}+\frac{1}{2}VS^{2}\frac{\partial
^{2}P_{1}}{\partial S^{2}}+\rho \eta SV\frac{\partial ^{2}P_{1}}{\partial
S\partial V}\right.  \\
&&\left. +\frac{1}{2}V\eta ^{2}\frac{\partial ^{2}P_{1}}{\partial V^{2}}+rS%
\frac{\partial P_{1}}{\partial S}-rV\right) /\frac{\partial P_{1}}{\partial V%
}
\end{eqnarray*}

Having obtained this equation one can conclude that both left and right-hand
sides should be equal to some function $g$ that only depends on the
independent variables $S,V$ and $t$. If one define the function $g$ to have
the following form $g=\kappa \left( V-\theta \right) -\lambda V$ a special
case of a so-called affine diffusion process a considered. For such class of
processes, the pricing PDE is tractable analytically.

In the considered case the Garman's pricing PDE for the Heston model is
given by Eq.(\ref{Heston-PDE})

\bigskip


\section{Appendix B}


In this appendix some results regarding the work of Duffie, Pan and
Singleton (2000) \cite{Duffie-Pan-Singleton}, will be summarized. For affine
diffusion processes the characteristic function of $x(T)$ can be written as
given in Eq.(\ref{characteristic function-1})%
\begin{equation*}
f(x,V,\tau ,\omega )=\exp \left[ A\left( \omega ,\tau \right) +B\left(
\omega ,\tau \right) V+i\omega x\right]
\end{equation*}%
The characteristic function $f(x,V,\tau ,\omega )$ satisfies the forward
equation given in Eq.(\ref{Heston-forwardPDE}). Substituting $f(x,V,\tau
,\omega )$ given in Eq.(\ref{characteristic function-1}) into Eq.(\ref%
{Heston-forwardPDE}) yields%

\begin{widetext}
\begin{equation}
-f\left( \frac{\partial A}{\partial \tau }+\frac{\partial B}{\partial \tau }%
V\right) +\frac{1}{2}Vf\left( -\omega ^{2}-i\omega \right) +\rho \eta
VBfi\omega +\frac{1}{2}\eta ^{2}VB^{2}f-\kappa (V-\theta )Bf=0
\end{equation}
\end{widetext}

If one use the notations given in Eq.(\ref{A-B-ODE-params}), than Eq.(\ref%
{AppB-characteristic eq}) simplifies to
\begin{equation*}
-\frac{\partial A}{\partial \tau }+aB+V\left( \frac{\partial B}{\partial
\tau }+\alpha -\beta B+\gamma B^{2}\right) =0
\end{equation*}

The equation above is a first order polynomial in $V.$ In order for this
equation to hold, both coeficients must vanish, i.e. we obtain that $A\left(
\omega ,\tau \right) $ and $\ B\left( \omega ,\tau \right) $ satisfy the
system of ordinary differential equations ODE given in Eq.(\ref{A-B-ODE}).


\section{Appendix C}


This appendix\ is written for the readers convenience and contains basic
definitions and formulas for the confluent hypergeometric function.

Standard form of the confluent hypergeometric equation reads%
\begin{equation}
z\frac{d^{2}w}{dz^{2}}+\left( b-z\right) \frac{dw}{dz}-aw=0
\label{AppC-hypergeometric-eq}
\end{equation}

This equation has a regular singularity at $z=0$ and an irregular
singularity at $z=\infty .$ There are eight linearly independent solutions,
but the standard two are so-called Kummer's Functions

\begin{equation}
M(a,b,z)=1+\frac{az}{b}+\frac{\left( a\right) _{2}z^{2}}{\left( b\right)
_{2}2!}+...+\frac{\left( a\right) _{n}z^{n}}{\left( b\right) _{n}n!}+...
\label{Kummer-M}
\end{equation}

where
\begin{equation*}
\left( a\right) _{n}=a(a+1)(a+2)...(a+n-1),\text{ \ \ }\left( a\right) _{0}=1
\end{equation*}

and the second function is given by%
\begin{eqnarray}
U(a,b,z) &=&\frac{\pi }{\sin \pi b}\left[ \frac{M(a,b,z)}{\Gamma
(1+a-b)\Gamma (b)}\right.  \label{Kummer-U} \\
&&\left. -z^{1-b}\frac{M(1+a-b,2-b,z)}{\Gamma (a)\Gamma (2-b)}\right]  \notag
\end{eqnarray}

Using the $M(a,b,z)$ and $U(a,b,z)$ the complete solution of the confluent
hypergeometric equation Eq.(\ref{AppC-hypergeometric-eq}) reads%
\begin{equation}
w(z)=AM(a,b,z)+BU(a,b,z)  \label{AppC-confluent eq-solution}
\end{equation}

where $A$ and $B$ are arbitrary constants and $b\neq -n$. Within the related
literature are used alternative notations for $M(a,b,z)$ and $U(a,b,z),$
which respectively reads $_{1}F_{1}(a;b;z)$ and $z^{-a}$ $%
_{2}F_{0}(a,1+a-b;;-1/z).$ The derivative of the two independent solutions
with respect to the independent variable $z$ are given by%
\begin{equation}
\frac{d}{dz}M(a,b,z)=\frac{a}{b}M(a+1,b+1,z)  \label{dz-M}
\end{equation}

and%
\begin{equation}
\frac{d}{dz}U(a,b,z)=-aU(a+1,b+1,z)  \label{dz-U}
\end{equation}

\bigskip


\section{Appendix D}


The appendix contains basic definitions and formulas for the Heun equation
and more precisely for the triconfluent Heun equation, which is one of the
confluent forms of Heun's equation. Heun's equation
\begin{equation}
\frac{d^{2}w}{dz^{2}}+\left( \frac{\gamma }{z}+\frac{\delta }{z-1}+\frac{%
\epsilon }{z-a}\right) \frac{dw}{dz}+\frac{\alpha \beta z}{z(z-1)(z-a)}w=0
\label{AppB-Heun}
\end{equation}

has regular singularities at $0,1,a$ and $\infty $. Generally speaking
confluent forms of Heun's differential equation arise when two or more of
the regular singularities merge to form an irregular singularity. This
process is analogous to

the derivation of the confluent hypergeometric equation from the
hypergeometric equation. There are four confluent standard forms and the
triconfluent Heun Equation given by%
\begin{equation}
\frac{d^{2}w}{dz^{2}}+(\gamma +z)z\frac{dw}{dz}+(\alpha z+q)w=0
\label{Triconfluent-eq}
\end{equation}

This equation has one singularity, which is an irregular singularity of rank
$3$ at $z=$ $\infty .$ Within the literature one can find other canonical
forms of the triconfluent Heun Equation. Following Slavyanov, S.Y., and Lay,
W the

HeunT$(\alpha ,\beta ,\gamma ,z)$ function is a local solution to Heun's
Triconfluent equation,

\begin{equation}
\frac{d^{2}w}{dz^{2}}-\left( 3z^{2}-a\right) \frac{dw}{dz}-\left( \left(
3-b\right) z-a\right) w=0  \label{Triconfluent-eq1}
\end{equation}

computed as a standard power series expansion around the origin, which is a
regular point. Because the single singularity is located at $z=$ $\infty $,
this series converges in the whole complex plane.



\end{document}